\documentclass[twocolumn,showpacs,preprintnumbers,amsmath,amssymb,prc]{revtex4}

\usepackage{graphicx}
\usepackage{dcolumn}
\usepackage{bm}

\usepackage{mathptmx, courier, pifont}
\usepackage[scaled=0.92]{helvet}
\usepackage[T1]{fontenc}
\usepackage{textcomp}

\newcommand{\quarterthin}{\kern 0.0417em}

\begin{document}


\title{Shell model study for neutron-rich $sd$-shell nuclei}

\author{Kazunari Kaneko$^{1}$, Yang Sun$^{2,3}$, Takahiro Mizusaki$^{4}$,
Munetake Hasegawa$^{2}$}

\affiliation{
$^{1}$Department of Physics, Kyushu Sangyo University, Fukuoka
813-8503, Japan \\
$^{2}$Institute of Modern Physics, Chinese Academy of Sciences,
Lanzhou 730000, People's Republic of China \\
$^{3}$Department of Physics, Shanghai Jiao Tong
University, Shanghai 200240, People's Republic of China \\
$^{4}$Institute of Natural Sciences, Senshu University, Tokyo
101-8425, Japan
}

\date{\today}

\begin{abstract}
The microscopic structure of neutron-rich $sd$-shell nuclei is
investigated by using the spherical shell-model in the $sd$-$pf$
valence space with the extended pairing plus quadrupole-quadrupole
forces accompanied by the monopole interaction (EPQQM).  The
calculation reproduces systematically the known energy levels for
even-even and odd-mass nuclei including the recent data for
$^{43}$S, $^{46}$S and $^{47}$Ar. In particular, the erosion of the
$N=28$ shell closure in $^{42}$Si can be explained.  Our EPQQM
results are compared with other shell model calculations with the
SDPF-NR and SDPF-U effective interactions.
\end{abstract}


\maketitle

\section{Introduction}\label{sec1}

The current experimental and theoretical investigation focuses on
the study of evolution of nuclear structure around the shell gap at
$N=$ 28. The $N=28$ shell closure is a traditional one in the
nuclear single-particle spectrum driven by the spin-orbit
interaction. For example, the $^{48}$Ca nucleus has been known to
exhibit double-magicity as a result of the neutron subshell gap
separating the $f_{7/2}$-orbit and the $f_{5/2}p$-shells at $N=28$.
However, recent theoretical studies and experimental data have
questioned the persistence of the traditional magic numbers, and
revealed that the $N=$ 28 shell gap is eroded when moving away from
the stability line. The experimental data for the $^{44}$S and
$^{42}$Si isotones indicate a clear breaking of the $N=$ 28 magicity
because of the observed small $2^{+}$ energy
\cite{Scheit,Sohler02,Bastin07}. Recently, a direct evidence of collapse of
the $N=$ 28 shell closure due to the level inversion between the
lowest 7/2$^{-}$ and 3/2$^{-}$ levels has been observed in $^{43}$S
\cite{Gaudefroy09}. In addition, toward $N=28$, the degeneracy
between the lowest $1/2^{+}$ and $3/2^{+}$ states in the odd-proton
P, Cl, and K isotopes suggests \cite{Sorlin04} that the shell gap
between the $d_{3/2}$ and $s_{1/2}$ proton orbits almost collapses
at $N=28$. The reduction of the $N=28$ neutron gap combined with
this degeneracy is regarded as the possible origin of erosion of the
$N=28$ magicity in the neutron-rich $sd$-shell nuclei.

To gain the insight for the evolution of shell structure in the
neutron-rich $sd$-shell nuclei and to understand the details of
collapse of the $N=28$ magicity, shell-model calculations in the
full $sd$-$pf$ space are desirable. However, conventional
shell-model calculations in the full $sd$-$pf$ space are not
possible at present because of the huge dimension in the
configuration space. To study the neutron-rich $sd$-shell nuclei,
therefore, one needs to incorporate truncations in the valence
space. One choice is such that valence protons are restricted to the
$sd$ shell and neutrons to the $sd$-$pf$ shell with twelve $sd$
frozen neutrons. Such an approach was originally introduced with the
effective interaction SDPF-NR \cite{Retamosa97,Nummela01,Caurier98,
Caurier02,Caurier05}. The nuclei in this mass region are known
to have a variety of shapes (e.g. shape coexistence discussed in Refs.
\cite{Sohler02,Gaudefroy09}), and the nucleus $^{42}$Si has been the
subject of debate on its magic nature \cite{Fridmann05,Bastin07}.
The SDPF-NR interaction was applied to the neutron-rich $sd$-shell
nuclei, and one found that it describes well the excitation energies
for very neutron-rich isotopes with $Z > 14$. However, the
calculation deviated from the experimental data for the silicon
($Z=14$) isotopes where the calculated energies for $^{36-42}$Si
were too high. To improve the results, the $pf$-shell effective
interaction in SDPF-NR was renormalized to compensate for the
absence of 2p-2h excitations from the core \cite{Nowacki}. With the
reduction of pairing in the renormalization for $Z\leq 14$, the
$2_{1}^{+}$ excitation energies in silicon isotopes agreed nicely
with experiment. Thus, two SDPF-U interactions, one for $Z> 14$ and
the other for $Z\leq 14$, are respectively needed for the
description of collectivity at $N=28$ in the isotopic chains of
sulfur ($Z=$16) and silicon ($Z=$14). In a similar way, the SDPF-NR2
and SDPF-NR3 interactions have recently been suggested to describe
the sudden change in nuclear structure at $N=28$ \cite{Gade09}. From
the above discussion, one sees that a consistent shell-model
treatment for collectivity at the $N=28$ shell closure in different
isotopic chains has not been successful so far.

The main purpose of the present article is to construct a unified
effective interaction for the neutron-rich $sd$-shell nuclei, which
can be consistently applicable to both $Z > 14$ and $Z\leq 14$
isotopic chains, and to understand the shell evolution in this
exotic mass region. It is well known that realistic effective
interactions used in the low-energy nuclear structure study are
dominated by the pairing plus quadrupole-quadrupole ($P+QQ$) forces
with inclusion of the monopole term \cite{Dufour}. As documented in
the literature, the extended $P+QQ$ model combined with the monopole
interaction works well for a wide range of nuclei
\cite{Hasegawa01,Kaneko02}. This effective interaction is called
hereafter EPQQM to distinguish it from shell models with other
effective interactions. The EPQQM model has demonstrated its
capability of describing the microscopic structure in different
$N\approx Z$ nuclei, as for instance, in the $fp$-shell region
\cite{Hasegawa01} and the $fpg$-shell region \cite{Kaneko02}.
Recently, it has been shown that the EPQQM model is also applicable
to the neutron-rich Cr isotopes \cite{Kaneko08}.

The monopole interaction plays an important role in our discussion.
The monopole shifts in the spherical shell model have been
introduced to account for the non-conventional shell evolution in
neutron-rich nuclei. Connection between the monopole shifts and the
tensor force \cite{Otsuka01} has been studied within the
self-consistent mean-field model using the Gogny force
\cite{Otsuka05}. It has recently been clarified that the
possible physical origin of these interactions is attributed to the
central and tensor forces \cite{Smirnova10,Otsuka10}.  For example,
when one goes from the Ca down to the Si isotopes a significant
reduction of the $N=28$ shell gap can be found. By removing protons
from the $d_{3/2}$ orbit in nuclei between Ca and Si, the strong
attractive proton-neutron monopole force between the $\pi d_{3/2}$
and $\nu f_{7/2}$ orbits is no longer present
\cite{Nowacki,Otsuka08}. This makes the neutron $f_{7/2}$ less
bound, and thus reduces the size of the $N=28$ gap in nuclei between
$^{40}$Ca and $^{34}$Si.

The paper is arranged as follows. In Sec.~\ref{sec2}, we outline our
model. In Section~\ref{sec3}, we perform the numerical calculations
and discuss the results for the neutron-rich nuclei in the $sd$-$pf$
shell region. Finally, conclusions are drawn in Section~\ref{sec4}.

\section{The model}\label{sec2}

\begin{figure}[b]
\includegraphics[totalheight=5.5cm]{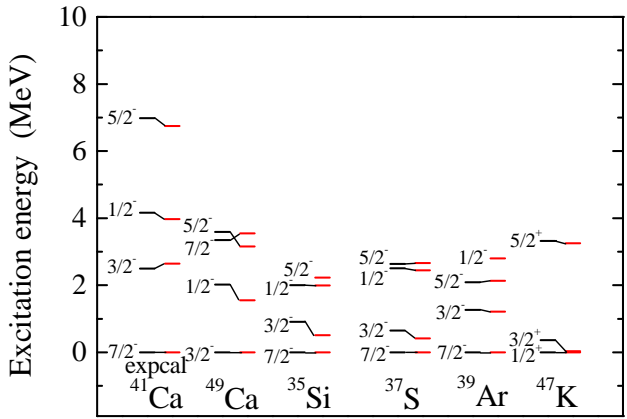}
  \caption{(Color online) Comparison of experimental and calculated energy levels
  for the odd-mass nuclei near the neutron or proton shell closure in the $sd$ region. }
  \label{fig1}
\end{figure}
\begin{figure}[t]
\includegraphics[totalheight=5.5cm]{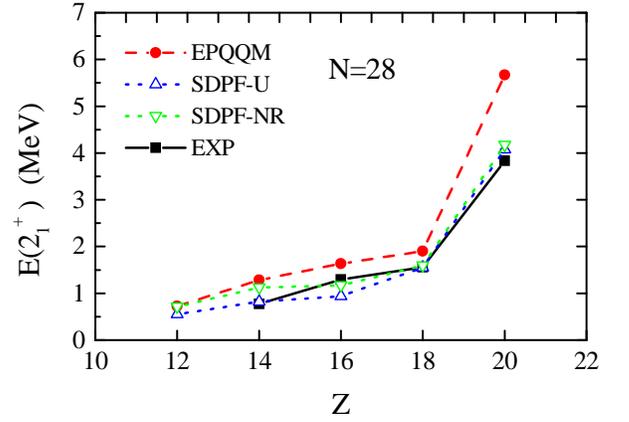}
  \caption{(Color online) Experimental and calculated first excited $2^{+}$ energies
  $E(2_{1}^{+})$ for the neutron-rich $N=28$ isotones.}
  \label{fig2}
\end{figure}

We start with the following form of Hamiltonian, which consists of
pairing and quadrupole-quadrupole terms with the monopole
interaction
\begin{eqnarray}
 H & = & H_{\rm sp} + H_{P_0} + H_{P_2} + H_{QQ} + H_{\rm m}  \nonumber \\
   & = & \sum_{\alpha} \varepsilon_a c_\alpha^\dag c_\alpha
    -  \sum_{J=0,2} \frac{1}{2} g_J \sum_{M\kappa} P^\dag_{JM1\kappa} P_{JM1\kappa}  \nonumber \\
   & - & \frac{1}{2} \chi_2/b^{4} \sum_M :Q^\dag_{2M} Q_{2M}:  \nonumber \\
   & + & \sum_{a \leq b} \sum_{T} k_{\rm m}^T(ab) \sum_{JMK}
               A^\dagger_{JMTK}(ab) A_{JMTK}(ab),
          \label{eq:0}
\end{eqnarray}
where $b$ in the third term is the length parameter of harmonic
oscillator. We take the $J=0$ and $J=2$ forces in the pairing
channel, and the quadrupole-quadrupole ($QQ$) forces in the
particle-hole channel \cite{Hasegawa01,Kaneko02}. The monopole
interaction is denoted by $H_{\rm m}$, where the global monopole
force is neglected because it does not affect the excitation
energies of the low-lying states. This isospin-invariant Hamiltonian
(\ref{eq:0}) is diagonalized in a chosen model space based on the
spherical basis. We employ the shell-model code ANTOINE
\cite{Antoine} for the numerical calculation. In the present work,
we consider an $^{16}$O core and employ the $sd$-$pf$ model space
comprising the $0d_{5/2}$, $1s_{1/2}$, $0d_{3/2}$ active proton
orbitals and the $0d_{5/2}$, $1s_{1/2}$, $0d_{3/2}$, $0f_{7/2}$,
$1p_{3/2}$, $0f_{5/2}$, $1p_{1/2}$ active neutron orbitals with
twelve $sd$ frozen neutrons. We employ the same single-particle
energies as those of Ref. \cite{Nowacki}: $\varepsilon_{d5/2} =0.0$,
$\varepsilon_{s1/2} =0.78$, $\varepsilon_{d3/2} =5.60$,
$\varepsilon_{f7/2} =9.92$, $\varepsilon_{p3/2} =10.01$,
$\varepsilon_{p1/2} =15.15$, and $\varepsilon_{f5/2} =10.18$ (all in
MeV). We adopt the following interaction strengths for the EPQQM
forces: $g_0 = 0.540, g_2 = 0.678$, and $\chi_2 = 0.474$ (all in
MeV). Finally, for the monopole terms, the strengths are chosen to
be (all in MeV):
\begin{eqnarray}
  & {} & k_{\rm m}^{T=0}(f_{7/2},d_{3/2}) = -0.65,
   \quad k_{\rm m}^{T=0}(f_{7/2},d_{5/2}) = -0.3,  \nonumber \\
  & {} & k_{\rm m}^{T=0}(d_{5/2},p_{1/2}) = 0.55,
   \quad k_{\rm m}^{T=0}(d_{5/2},p_{3/2}) = -0.2,  \nonumber \\
  & {} & k_{\rm m}^{T=0}(d_{3/2},p_{3/2}) = 0.5,
   \quad k_{\rm m}^{T=0}(s_{1/2},p_{3/2}) = -0.6,  \nonumber \\
  & {} & k_{\rm m}^{T=1}(d_{5/2},d_{3/2}) = -0.2,
   \quad k_{\rm m}^{T=1}(f_{7/2},f_{5/2}) = -0.1,  \nonumber \\
  & {} & k_{\rm m}^{T=1}(f_{7/2},p_{1/2}) = -0.01,
   \quad k_{\rm m}^{T=1}(f_{7/2},p_{3/2}) = -0.03, \nonumber \\
  & {} & k_{\rm m}^{T=1}(f_{7/2},f_{7/2}) = -0.3,
   \quad k_{\rm m}^{T=1}(p_{3/2},p_{3/2}) = -0.45.
\label{eq:2}
\end{eqnarray}
The rest of the monopole terms are neglected in the present
calculations. The above EPQQM force strengths are determined so as
to reproduce the experimental energy levels for odd-mass nuclei that
have one particle or one hole on top of the neutron and proton shell
closures. The results are compared with experimental data
\cite{ensdf} in Fig. \ref{fig1}. Here the root mean square
(RMS) deviations between the experimental and theoretical excitation
energies are 0.24, 0.14, and 0.20 (in MeV) for the $3/2^{-}$,
$1/2^{-}$, and $5/2^{-}$ states, respectively, which means that the
agreement is good. We emphasize that in order to describe the data
correctly, a very important part of the interaction is those
monopole forces. The attractive $T=0$ monopole interactions
$\pi d_{3/2}$-$\nu f_{7/2}$ and $\pi d_{5/2}$-$\nu f_{7/2}$ have
recently been proposed to be attributed to the monopole effect of
the central and tensor forces in the nucleon-nucleon interaction
\cite{Smirnova10,Otsuka10}, where the central term produces a global
contribution and the tensor term generates local variations.  They
are responsible for the collapse of the $N=28$ shell closure in the
neutron-rich $sd$-shell nuclei, which gives rise to the drastic
change in nuclear shape of the $N=28$ isotones. In fact, it has been
reported that the tensor force produces the pronounced oblate
minimum in the potential energy surface for $^{42}$Si
\cite{Otsuka08}. Along the K isotopic chains, the repulsive $\pi
d_{3/2}$-$\nu p_{3/2}$ and attractive $\pi s_{1/2}$-$\nu p_{3/2}$
monopole interactions lead to an inversion of the expected ordering
in the lowest $3/2^{+}$ and $1/2^{+}$ levels in $^{47}$K. In
addition, these interactions reproduce the evolution of the lowest
$3/2^{-}$ state in $^{35}$Si. The monopole interactions between $\pi
d_{5/2}$ and the $pf$-shells act so as to reproduce the low-lying
states in $^{49}$Ca.

\begin{figure}[t]
\includegraphics[totalheight=5.5cm]{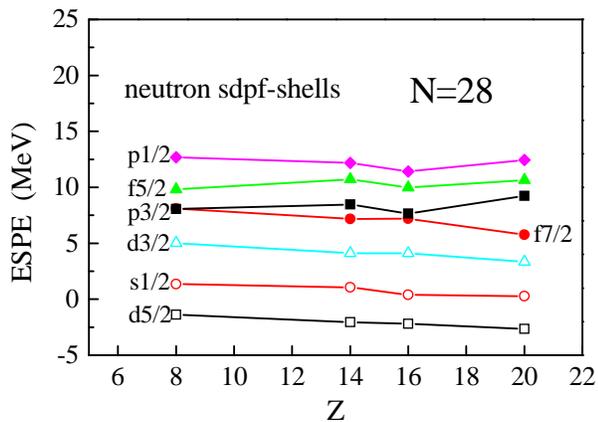}
  \caption{(Color online) Effective neutron single-particle energies at $N=28$
from $Z=8$ to $Z=20$.}
  \label{fig3}
\end{figure}
\begin{figure}[t]
\includegraphics[totalheight=5.5cm]{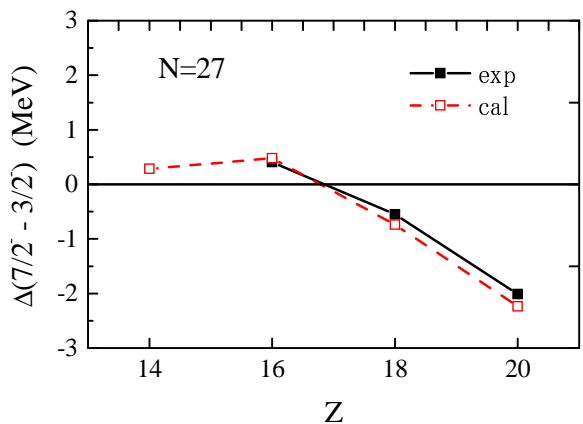}
  \caption{(Color online) Energy splitting between the lowest $7/2^{-}$ and
$3/2^{-}$ states in neutron-rich $N=27$ isotones \cite{Gaudefroy09,Gaudefroy08,Riley08}.}
  \label{fig4}
\end{figure}

\section{Numerical results and discussions}\label{sec3}

\subsection{The shell erosion in $N=28$}

We discuss changes in the shell structure when moving away from the
valley of stability. Large energy gaps at the traditional magic
numbers in stable nuclei may be washed out in the neutron- or
proton-rich regions. The fundamental question of how the nuclear
shell structure evolves with proton or neutron excess is one of the
main motivations to study nuclei far from stability. The study of
shapes in such exotic nuclei serves as a sensitive test for the
predictive power of nuclear models. Figure \ref{fig2} shows the
systematics of the first excited $2^{+}$ states for the $N=28$
isotones, which are compared with three shell model calculations,
respectively with the EPQQM, SDPF-U, and SDPF-NR effective
interactions. As one can see, all the calculations are capable of
producing the drastic drop in energy when removing protons from the
double magic nucleus $^{48}$Ca.

To understand the erosion of the $N=28$ shell gap, the neutron
effective single-particle energies (ESPE) as a function of proton
number are shown in Fig. \ref{fig3}. 
The $\nu f_{7/2}$ and $\nu p_{3/2}$ orbitals at $Z=20$ becomes
degenerate at $Z=16$.
This degeneracy is due to the attractive proton-neutron monopole
interaction between the $\nu f_{7/2}$ and $\pi d_{3/2}$ orbitals. As
seen in Fig. \ref{fig4}, the inversion of the lowest $7/2^{-}$ and
$3/2^{-}$ states in $^{43}$S ($Z=16$ and $N=27$) just reflects the
degeneracy of the two orbitals, $\nu f_{7/2}$ and $\nu p_{3/2}$, at
$Z=16$ in Fig. \ref{fig3}. 
On the other hand, the neutron ESPE in the SDPF-U interaction
does not show such a strong reduction of the $N=28$ shell gap from
$Z=20$ to $Z=16$. At $Z= 16$ the gap still equals about 3.5 MeV, and
the inversion between the $7/2^{-}$ and $3/2^{-}$ states is ascribed
to the combined effect of the gap reduction due to the monopole
interactions and the increase of the multipole correlations.
The $N=28$ shell closure is eroded in
$^{46}$Ar and $^{44}$S, after the removal of only two and four
protons, respectively. This rapid disappearance of rigidity in the
$N=28$ isotones has been ascribed to a reduction of the neutron
shell gap at $N=28$ combined with that of the proton subshell gap at
$Z=16$, leading to increased probability of quadrupole excitations
within the $fp$ and $sd$ shells for neutrons and protons,
respectively. In fact, the occurrence of the low-lying isomer at
320.5 keV in $^{43}$S has been interpreted in the shell-model
framework as resulting from the inversion between $(\nu
7/2^{-})^{-1}$ and $(\nu 3/2^{-})^{+1}$ configurations
\cite{Gaudefroy08,Sarazin00}. We have thus understood that the
inversion has the origin of vanishing $N=28$ magicity at $Z=16$.

\begin{figure}[t]
\includegraphics[totalheight=5.5cm]{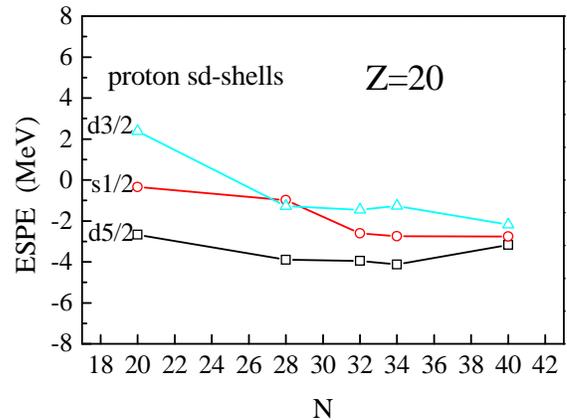}
  \caption{(Color online) Effective proton single-particle energies at $Z=20$
from $N=20$ to $N=40$.}
  \label{fig5}
\end{figure}

\begin{figure}[b]
\includegraphics[totalheight=6.5cm]{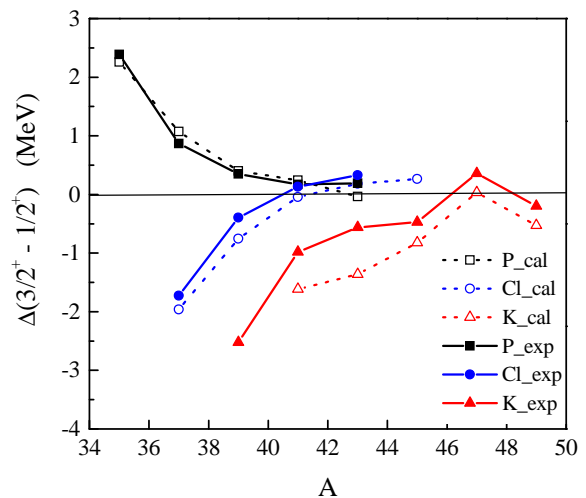}
  \caption{(Color online) Energy splitting between the lowest $3/2^{+}$ and
$1/2^{+}$ states in neutron-rich P, Cl, and K isotopes.}
  \label{fig6}
\end{figure}

In Fig. \ref{fig5}, the proton ESPE with $Z=20$ are shown as a
function of neutron number. It is seen that the shell gap between
the $\pi d_{3/2}$ and $\pi s_{1/2}$ proton orbitals in the double
closed-shell nucleus $^{40}$Ca ($Z=N=20$) decreases with increasing
neutron number. The $\pi d_{3/2}$ orbit is almost degenerate with
the $\pi s_{1/2}$ orbit at $N=28$, and lies just below the $\pi
s_{1/2}$ orbit at $N=28$. An inversion of the $\pi d_{3/2}$ and $\pi
s_{1/2}$ orbits thus occurs above $N=32$. The effect of adding
neutrons to the $f_{7/2}$ orbital is primarily to reduce the proton
$s_{1/2}$-$d_{3/2}$ gap. The relevant orbitals, $\pi s_{1/2}$ and
$\pi d_{3/2}$, are known to become degenerate at $N=28$. This
degeneracy enhances the quadrupole correlation energy of the
configurations with open neutron orbitals. 
The ESPE from $N=34$ to $N=40$ for the $\pi d_{5/2}$ are different
from those of the SDPF-U calculations. It would be attributed to
the monopole effect between the $\pi d_{5/2}$ and the $\nu f_{5/2}$.
Figure \ref{fig6} shows
the energy splitting between the lowest $3/2^{+}$ and $1/2^{+}$
states as a function of mass number for the neutron-rich isotopes of
phosphorus, chlorine, and potassium. The calculations are in a good
agreement with available experimental data
\cite{ensdf,Bastin07,Riley09} for the phosphorus and chlorine
isotopes, but not for the potassium isotopes.
The energy splitting between $3/2^{+}$ and $1/2^{+}$ for K isotopes
in Fig. \ref{fig6} deviates from the experimental data. This would be
related with the disagreement of the first excited $2^{+}$ enegies in
Fig. \ref{fig2}.
The variation trend
reflects the reduction of the $\pi d_{3/2}$-$\pi s_{1/2}$ splitting
in Fig. \ref{fig5} as neutron number increases. The inversion of the
$\pi d_{3/2}$-$\pi s_{1/2}$ orbits at $N=28$ corresponds to the
observation that the ground-state of $^{47}$K is actually the
$1/2^{+}$ state, not the expected $3/2^{+}$ from the hole state
$(\pi d_{3/2})^{-1}$. The near degeneracy of $\pi d_{3/2}$ and $\pi
s_{1/2}$ is again attributed to the attractive proton-neutron
monopole interaction between the $\nu f_{7/2}$ and $\pi d_{3/2}$.
The erosion of the $N=28$ shell gap is enhanced at $Z=16$ by the
degeneracy of the $\pi d_{3/2}$ and $\pi s_{1/2}$ proton orbits. In
Fig. \ref{fig6}, we also show the energy splitting between the
lowest $3/2^{+}$ and $1/2^{+}$ states for the $^{37-45}$Cl nuclei
\cite{Riley09}.

\subsection{Comparison with other shell model calculations}

A consistent description of collectivity at $N=28$ for different
isotopic chains of the neutron-rich $sd$-shell nuclei has been a
challenge for shell model calculations. The early calculations with
the SDPF-NR interaction described the level scheme and transitions
for these isotopic chains. However, it was found that the calculated
results deviate from the experimental data. For the silicon isotopes
the calculated energy levels are too high. To improve the agreement,
Nowacki and Poves have proposed two effective interactions
\cite{Nowacki}, one (SDPF-U) for $Z \leq 14$ and the other (SDPF-U1)
for $Z > 14$.

\begin{figure*}[t]
\includegraphics[totalheight=8.5cm]{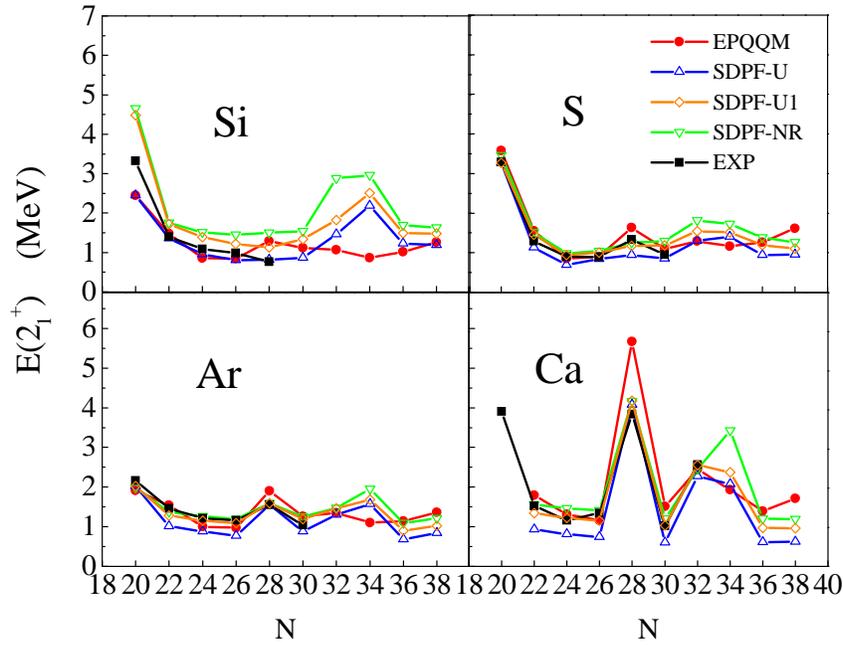}
\caption{(Color online) Comparison between experimental and
calculated first excited $2^{+}$ energies $E(2_{1}^{+})$ with several
effective interactions, EPQQM, SDPF-U, SDPF-U1, and SDPF-NR, for Si
$(Z=14)$, S $(Z=16)$, Ar $(Z=18)$, and Ca $(Z=20)$ isotopes. }
  \label{fig7}
\end{figure*}

It is interesting to compare the energy levels obtained from the
shell model calculations with different effective interactions
(EPQQM, SDPF-U, SDPF-U1, and SDPF-NR). Figure \ref{fig7} shows the
first excited $2^{+}$ energies as a function of neutron number for
the Ca, Ar, S, and Si isotopes. The EPQQM calculations agree well
with the known experimental data, except for $^{34}$Si and $^{44}$S.
The SDPF-NR calculations also agree with the data for the Ca and Ar
isotopes, but cannot reproduce those of the Si isotopes. The
calculated first excited $2^{+}$ states lie higher than experiment.
On the other hand, the SDPF-U (SDPF-U1) interaction fails to
reproduce the known experimental data for $Z > 14$ ($Z \leq 14$).
Moreover, considerable differences between the EPQQM and SDPF-NR
(and also SDPF-U, SDPF-U1) can be seen around the neutron number
$N=$ 34, where no experimental data are currently available to
discriminate the predictions. We note that none of the interactions
can reproduce the first excited $2^{+}$ energy level of $^{34}$Si with
$N=20$. This suggests that it would be necessary to include the neutron
excitations from the $sd$-shell to the $pf$-shell across the $N=20$
energy gap. The disappearance of the $N=20$ magic structure between
the $sd$ and $pf$ shells has already been established for the
neutron-rich nuclei with $N=$ 20 isotones.

\begin{figure}[t]
\includegraphics[totalheight=5.5cm]{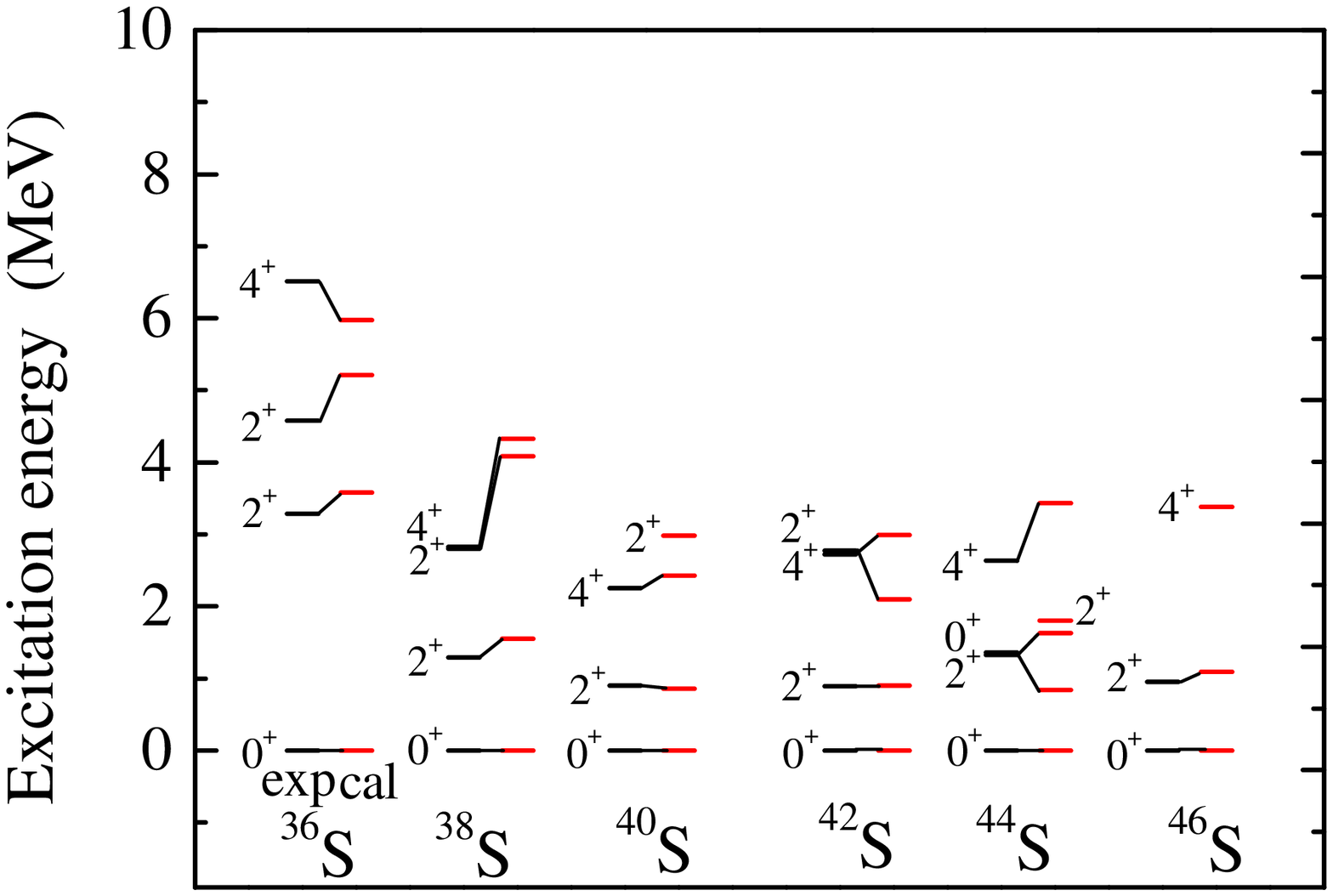}
  \caption{(Color online)Experimental and calculated energy levels for
even-even Sulfur isotpes
\cite{ensdf,Sohler02,Stuchbery06,Force10}. }
  \label{fig8}
\end{figure}

\begin{figure}[b]
\includegraphics[totalheight=5.5cm]{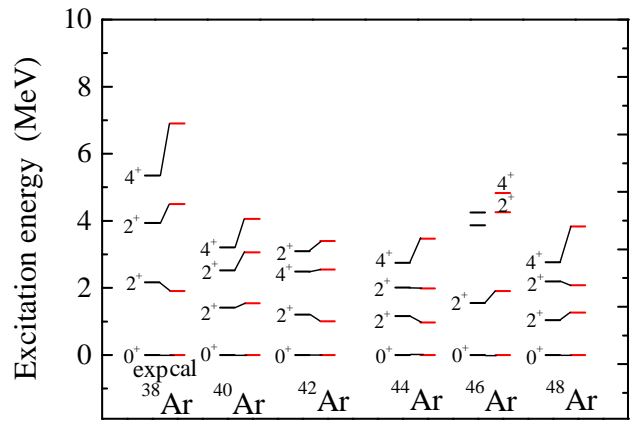}
  \caption{(Color online)Experimental and calculated energy levels for
even-even Argon isotopes
\cite{ensdf,Gade03,Riley05,Zielinska09}. }
  \label{fig9}
\end{figure}

\subsection{Level scheme and $B(E2)$ values}

In this subsection, we present the theoretical level schemes and
electromagnetic $E2$ transition probabilities calculated with the
EPQQM interaction, and compare the results with experiment. In Fig.
\ref{fig8}, we show energy levels for the even-even sulfur isotopes,
where for each isotope, the left and right part of energy levels
denote the experimental data \cite{ensdf,Gade09,Force10} and the
calculated energies, respectively. The agreement between the
calculation and experiment is fairly good. In particular, the
calculated first excited $2^{+}$ energy level for $^{46}$S
reproduces very well the recently observed data \cite{Gade09}. Very
recently, energy spectrum for $^{44}$Ar has been experimentally
obtained \cite{Bhattacharyya08,Zielinska09}. In $^{40}$S, the
calculated $E(4_{1}^{+})/E(2_{1}^{+})$ ratio is about 2.5, which
qualitatively suggests a transitional or $\gamma$-soft nature for
this nucleus. In $^{42}$S, the $E(4_{1}^{+})/E(2_{1}^{+})$ ratio is
3.0, which is quite close to the rigid rotor value. The situation
significantly differs in $^{44}$S where both deformed and spherical
configurations are predicted to coexist and mix weakly with each other,
which is supported by the observation of the low-lying $0_{2}^{+}$
state at 1.326 MeV \cite{Force10}. In the shell model calculation, this state lies
below the $2_{1}^{+}$ state. The presence of a low-lying $0_{2}^{+}$
state is considered as a signature of a spherical-deformed
shape-coexistence and the present data support the weakening of the
$N=28$ shell gap. Figure \ref{fig9} shows the energy levels for
the even-even argon isotopes. One finds that the calculated energy
levels can reproduce the experimental data \cite{ensdf,Zielinska09}.
The first excited $2^{+}$ energy $E(2_{1}^{+})$ for Ar isotopes
decreases with increasing neutron number from $^{38}$Ar to
$^{44}$Ar, but with a sudden increase for the $N=28$ isotope. 
This suggests the persistence of the $N=28$ gap in Ar nuclei.
In Figs. \ref{fig8}
and \ref{fig9}, we note that the RMS deviations for the $2_{1}^{+}$,
$2_{2}^{+}$, and $4_{1}^{+}$ states are respectively 0.22, 0.71, and
0.92 (in MeV), and the agreement becomes worse with increasing
excitation energy. 
The disagreement may require the further improvement of the $EPQQM$
interaction.

\begin{figure}[t]
\includegraphics[totalheight=4.5cm]{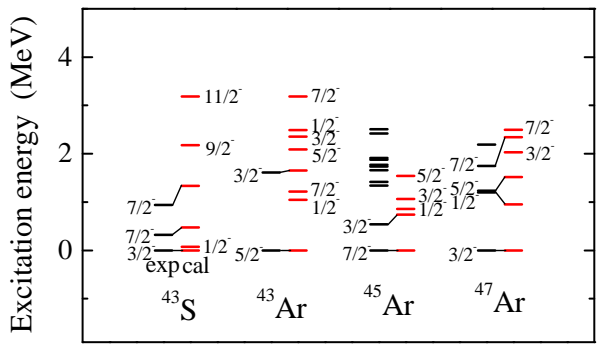}
  \caption{(Color online)Comparison of calculated energy levels with the
  recent experimental data for S and Ar nuclei
\cite{ensdf,Baumann07,Gaudefroy06,Gaudefroy08,Gaudefroy09,Bhattacharyya08,Gade05}. }
  \label{fig10}
\end{figure}

To demonstrate the EPQQM model calculation for odd-mass nuclei in
this mass region, we show in Fig. \ref{fig10} energy levels for some
odd-mass S and Ar isotopes. In the comparison, the experimental
energy levels for $^{43}$S and $^{47}$Ar are taken from the recent
experiments \cite{Gaudefroy09,Bhattacharyya08}. From the
calculation, the ground state of $^{43}$S is $3/2^{-}$, not the
expected $7/2^{-}$, suggesting that this nucleus is deformed. The
occurrence of a low-lying isomer at 320.5 keV in $^{43}$S is
interpreted in the shell-model framework as resulting from the
inversion between the natural $(\nu f_{7/2})^{-1}$ and the intruder
$(\nu p_{3/2})^{+1}$ configurations. Both the low
$B(E2;7/2^{-}\rightarrow 3/2^{-})$ value and the absence of
calculated deformed structure built on the $7/2^{-}$ isomer suggest
a coexistence of different shapes in the low-lying structure of the
nucleus $^{43}$S. In $^{47}$Ar, the ground state $3/2^{-}$ and the
excited state $1/2^{-}$ can be interpreted as a coupling of a
$p_{3/2}$ and $p_{1/2}$ neutron to the $0^{+}$ ground
state of $^{46}$Ar. The $5/2^{-}$ and $7/2^{-}$ levels result from
the coupling of a $p_{3/2}$ neutron to the $2_{1}^{+}$ state of
$^{46}$Ar. 

In Table I, we compare $B(E2)$ values between the theoretical
calculations and experiments for the S, Ar, and Si isotopes. The
experimental $B(E2)$ values are taken from Refs.
\cite{Stuchbery06,Scheit,Raman,Glasmacher}. 
In our $B(E2)$ calculations, we take the effective charges $e_\pi$
for protons and $e_\nu$ for neutrons as follows; $e_\pi=1.35e$ and 
$e_\nu=0.35e$ for S isotopes, $e_\pi=1.50e$ and $e_\nu=0.50e$ for
Ar isotopes, and $e_\pi=1.15e$ and $e_\nu=0.15e$ for Si isotopes.
On the other hand, in the SDPF-NR and SDPF-U calculations the effective
charges $e_\pi=1.35e$ and $e_\nu=0.35e$ suggested in Ref. \cite{Nowacki}
are used for S and Si isotopes, and the effective charges $e_\pi=1.50e$
and $e_\nu=0.50e$ are adopted for Ar isotopes. 
The $B(E2)$ values were
previously calculated and discussed with the SDPF-NR interaction by
Retamosa {\it et al.} \cite{Retamosa97}.  As one can see, our
EPQQM calculation is in a good agreement with the experimental
$B(E2)$ values for the $^{36-42}$S, but larger than the experimental
value for $^{44}$S. On the other hand, both the SDPF-U and SDPF-NR
calculations also reproduce the data $B(E2)$ except for $^{38,44}$S.
In Table I, we also compare the calculated $B(E2)$ values with data
for the Ar isotopes. The experimental $B(E2)$ values are taken from
Refs. \cite{ensdf,Stuchbery06,Raman,Speidel}. One observes that the
experimental $B(E2)$ increases monotonically from $^{38}$Ar to the
midshell nucleus $^{42}$Ar, and then decreases towards $^{46}$Ar. We
note that all the theoretical calculations agree fairly well for
$^{38-44}$Ar, but for $^{46}$Ar the calculated $B(E2)$ values are
all too large \cite{Robinson09,Riley05}. All the current shell
model calculations with different interactions yield a similar
value, which may indicate that this is a robust result. Very
recently, however, a new experimental value 114 $e^{2}$fm$^{4}$ has
been reported \cite{Mengoni10}, and is in a good agreement with all
the shell-model predictions. In Table I, this new experimental value
is included for comparison. It is apparent from the reduction in
excitation energy of the $2_{1}^{+}$ states and the enhancement in
$B(E2)$ that collective features are gradually developed in
neutron-rich isotopes between $N=20$ and $N=28$. However, a
different approach, the self-consistent mean-field calculation by
Werner {\it et al.} \cite{Werner94}, yields a much smaller $B(E2)$,
in a disagreement with the new experimental data. We compare
$B(E2)$ values between theoretical calculations and experiment for
the Si isotopes. The agreements with the available data are satisfactory.
The $B(E2)$ values increase with increasing neutron number $N$. The
calculated $B(E2)$ values are quite similar for the EPQQM and SDPF-U
interactions. For SDPF-NR, however, the predicted values are smaller
for the heavier isotopes $^{40,42}$Si.

\begin{table}[t]
\caption{Calculated $B(E2:2^{+}\rightarrow 0^{+})$ values for
         the even-even S, Ar, and Si isotopes.}
\begin{tabular*}{85mm}{@{\extracolsep{\fill}}cccccc} \hline\hline
        & \multicolumn{2}{c}{\hspace{2cm}$B(E2)$  [$e^2$fm$^{4}$]}    \\ \hline
    & Exp. & EPQQM & SDPF-U  & SDPF-NR  \\ \hline
$^{36}$S       &  21   &  22   &  22   &  21   \\
$^{38}$S       &  47   &  38   &  33   &  34   \\
$^{40}$S       &  67   &  63   &  58   &  60   \\
$^{42}$S       &  79   &  72   &  75   &  73   \\
$^{44}$S       &  63   &  98   &  73   &  73   \\
$^{38}$Ar      &  26   &  35   &  34   &  34   \\
$^{40}$Ar      &  66   &  59   &  49   &  50   \\
$^{42}$Ar      &  86   &  59   &  70   &  72   \\
$^{44}$Ar      &  69   &  47   &  71   &  77   \\
$^{46}$Ar      &  114  &  108  &  105  &  106  \\
$^{36}$Si      &  39   &  33   &  34   &  30   \\
$^{38}$Si      &  38   &  40   &  39   &  37   \\
$^{40}$Si      &   -   &  48   &  54   &  46   \\
$^{42}$Si      &   -   &  64   &  86   &  53   \\ \hline\hline
\end{tabular*}
\label{table1}
\end{table}

\begin{table}[t]
\caption{Calculated spectroscopic quadrupole moments for
         the first excited $2^{+}$ states in the $N=$28 isotones.}
\begin{tabular*}{85mm}{@{\extracolsep{\fill}}cccccc} \hline\hline
        & \multicolumn{2}{c}{\hspace{2cm}$Q$  [$e$fm$^{2}$]}    \\ \hline
    & Exp. & EPQQM & SDPF-U  & SDPF-NR  \\ \hline
$^{40}$Mg       &  -    &  -23   &  -20   &  -18   \\
$^{42}$Si       &  -    &   16   &   20   &   15   \\
$^{44}$S        &  -    &   -3   &  -16   &  -15   \\
$^{44}$Ar       &  -8   &  -16   &  -3.7  &  -2.5  \\
$^{46}$Ar       &  -    &   21   &   21   &   21   \\
$^{48}$Ca       &  -    &    5   &    3   &    3   \\ \hline\hline
\end{tabular*}
\label{table4}
\end{table}

Finally in Table II, we present calculated spectroscopic quadrupole
moments for the $N=$ 28 isotones. For all the effective
interactions, small, positive $Q$'s are found for $^{48}$Ca,
indicating a spherical character of this nucleus. For the
$^{44}$S nucleus, its small $2_{1}^{+}$ energy, large $B(E2)$ value,
and the presence of a second excited $0_{2}^{+}$ isomer at low
excitation suggest a mixed ground state configuration with spherical
and deformed shapes. Recent experiments on the odd mass sulfur
isotopes around $^{44}$S seem to indicate that $^{44}$S is a
deformed nucleus with strong shape coexistence whereas $^{42}$S can
be considered as a well-deformed system. The enhancement in $B(E2)$
of $^{46}$Ar coincides with a shape change from a small prolate
deformation in $^{44}$Ar to a large deformation in $^{46}$Ar. The
predicted spectroscopic quadrupole moment $Q_{s}(2_{1}^{+})$ in
$^{44}$Ar is somewhat larger than the experimental value, but with a
correct sign. In $^{42}$Si, the yrast sequence of $0_{1}^{+}$,
$2_{1}^{+}$, $4_{1}^{+}$ does not follow the rotational $J(J+1)$
law. However, its quadrupole moments are consistent with a deformed
oblate structure. The ground state of $^{40}$Mg and that of
$^{42}$Si are predicted to be prolately and oblately deformed,
respectively. All the shell model calculations indicate consistently
a rapid variation in quadrupole moment for the $N=28$ isotones,
suggesting a drastic shape change in this mass region.

\section{Conclusions}\label{sec4}

We have investigated the microscopic structure of the neutron-rich
$sd$-shell nuclei using the spherical shell model with a new
effective interaction in the $sd$-$pf$ valence space. The
interaction is of the extended pairing plus quadrupole-quadrupole
type forces with inclusion of the monopole interaction. The calculation
has reproduced reasonably well the known energy levels for even-even
and odd-mass nuclei including the recent data for $^{43}$S, $^{46}$S
and $^{47}$Ar. A special attention has been paid to the $N=28$
shell-gap erosion. It has been shown that the attractive $T=$ 0
monopole interactions $\pi d_{3/2}$-$\nu f_{7/2}$ and $\pi
d_{5/2}$-$\nu f_{7/2}$ are very important for the variation of the
shell gap at $N=28$. This shell evolution is considered as the
monopole effect of the central and tensor forces in the nucleon-nucleon
interaction. The strong attractive monopole interaction $\pi
d_{3/2}$-$\nu f_{7/2}$ causes the inversion of the $3/2^{+}$ and
$1/2^{+}$ states in odd-proton nucleus $^{47}$K. However, the
inversion is recovered in the neighboring nucleus $^{49}$K, which is
due to the repulsive monopole interaction $\pi d_{3/2}$-$\nu
p_{3/2}$. We have studied the shape change in the neutron-rich
$N=$28 isotones, and predicted the drastic shape change when protons
are removed away from the $Z=20$ isotone $^{48}$Ca, i.e. going
towards the neutron drip-line.

Research at SJTU was supported by the Shanghai Pu-Jiang grant, by
the National Natural Science Foundation of China under contract No.
10875077, and by the Chinese Major State Basic Research Development
Program through grant 2007CB815005.



\end{document}